\documentclass{ws-mpla}
\usepackage[super]{cite}
\usepackage{graphicx}
\begin{document}

\markboth{Naha Nzoupe, Dikand\'e, Tchawoua}
{Oscillons in parametrized $\phi^4$ models}

\catchline{}{}{}{}{}

\title{Kink-antikink scattering-induced breathing bound states and oscillons in a parametrized $\phi^4$ model
}

\author{F. Naha Nzoupe}
\address{Laboratory of Mechanics, Department of
Physics, Faculty of Science, University of Yaound\'e I
P.O. Box 812 Yaound\'e, Cameroon\\
}

\author{Alain M. Dikand\'e}
\address{Laboratory of Research on Advanced Materials and Nonlinear Science (LaRAMaNS), Department of Physics, Faculty of Science, University of Buea P.O. Box 63 Buea, Cameroon.\\
dikande.alain@ubuea.cm.}

\author{C. Tchawoua}

\address{Laboratory of Mechanics, Department of
Physics, Faculty of Science, University of Yaound\'e I
P.O. Box 812 Yaound\'e, Cameroon\\
}

\maketitle

\pub{Received (Day Month Year)}{Revised (Day Month Year)}

\begin{abstract}
Recent studies have emphasized the important role that a shape deformability of scalar-field models pertaining to the same class with the standard $\phi^4$ field, can play in controlling the production of a specific type of breathing bound states so-called oscillons. In the context of cosmology, the built-in mechanism of oscillons suggests that they can affect the standard picture of scalar ultra-light dark matter. In the present work kink scatterings are investigated in a parametrized model of bistable system admitting the classical $\phi^4$ field as an asymptotic limit, with focus on the formation of long-lived low-amplitude almost harmonic oscillations of the scalar field around a vacuum. The parametrized model is characterized by a double-well potential with a shape-deformation parameter that changes only the steepness of the potential walls, and hence the flatness of the hump of the potential barrier, leaving unaffected the two degenerate minima and the barrier height. It is found that the variation of the deformability parameter promotes several additional vibrational modes in the kink-phonon scattering potential, leading to suppression of the two-bounce windows in kink-antikink scatterings and the production of oscillons. Numerical results suggest that the anharmonicity of the potential barrier, characterized by a flat barrier hump, is the main determinant factor for the production of oscillons in double-well systems.   

\keywords{scalar field; parametrized $\phi^4$ model; instantons; kink-antikink collision; oscillons.}
\end{abstract}

\ccode{PACS Nos.: 03.50.-z; 05.45.Yv; 11.10.St; 03.65.Nk.}

\section{Introduction}
\label{intro}
The generation and interactions of solitary waves and solitons have attracted a great deal of interest over the past years, due to the fact that they can control many features related to the dynamics of natural systems ranging from biology and organic polymers, to classical and quantized fields in condensed-matter and high-energy physics \cite{1,1a,1b,1c,1d,zong}. The simplest localized solutions known in field theory are kink and antikink solitons, they display topological profiles in ($1+1$) space-time dimensions and can be generated in classical as well as quantum scalar field systems. \par 
In non-integrable scalar field theories such as the $\phi^{4}$ field \cite{1c,kv}, scatterings of a kink-antikink pair usually give rise to a competition between a bion state and a two-soliton solution characterized by a fractal structure in the parameter space of scattering velocity \cite{1e}. For some impact velocities the kink-antikink collision will give birth to a breather-like bound-state (bion) solution, that radiates progressively until a total annihilation of the pair. For other ranges of velocities, the pair performs an inelastic scattering with the solitons colliding once and separating thereafter. There also exist particular regions in velocity ($n-$bounce windows), where the scalar field at the center of mass can bounce several times ($n$ times) before the final separation of the pair. The later $n-$bounce windows have been explained as the consequence of a resonance mechanism for the exchange of energy between the vibrational and translational modes, resulting from discrete eigenstates of the Schr\"odinger-like equation inherent to the stability analysis of the $\phi^{4}$ kink \cite{1e,2}.\par
The last decade has witnessed a regain of interest in kink scatterings in non-integrable models, marked by intensive studies for instance of multi-kink collisions\cite{3,4,5,6,6a,29}, the interactions of a kink or an anti-kink with a boundary or a defect\cite{7,8}, the scattering processes in models with generalized dynamics \cite{9}, nonpolynomial models \cite{10,11,12,13}, polynomial models with one \cite{13a,14,7,16,17,18,19,20,21,22,23} and two \cite{24,25,26,27,27a} scalar fields and so on. However all these studies involve mostly two universal models which are the sine-Gordon model \cite{1}, assumed to describe systems with periodic one-site potentials, and the $\phi^{4}$ model intended for physical systems with double-well (DW) potentials. Although the $\phi^4$ kink for example has very recently been linked with topological excitations observed in buckled graphene nanoribbon \cite{28}, real physical systems to which the two universal models address are actually rather quite diverse, and most often unique in some aspects of their physical features. Indeed the $\phi^{4}$ and sine-Gordon model have fixed extrema while their shape profiles, including their their potential barriers, are rigid which confine their applicability to a very narrow class of physical systems. To lift the shortcomings related to the rigidity of shape profiles, these two universal models have been parametrized leading to two hierarchies of deformable-shape one-site potentials i.e. the Remoissenet-Peyard periodic potential \cite{remois1,remois,rem}, and the family of Dikand\'e-Kofan\'e (DK) DW potentials \cite{rem,dik1,dik2}. \par
In two recent studies \cite{13,31}, Bazeia {\it et al} addressed the issue of the influence of shape deformability of DW potentials, on kink-antikink scatterings with production of oscillon bound states \cite{6a,13,31}. Thus they first applied the shape deformability procedure to the standard $\phi^{4}$ by introducing a bistable model with non-polynomial potential, which they called sinh-deformed $\phi^{4}$ potential \cite{13}. Despite this new model showing similar features with the $\phi^{4}$ model a new phenomenon was observed, indeed under certain conditions the kink-antikink pair in the new model was found to convert, after collision, into long-lived low amplitude and almost harmonic oscillations of the scalar field around one vacuum. They interpreted these almost harmonic oscillations as a bound state of individual oscillons \cite{osc}. Later on the authors investigated \cite{31} kink-antikink collisions with production of oscillons, considering two members of the family of DK DW potentials \cite{dik1,dik2,32,33}. One member was a DW potential with variable separation between the two degenerate minima but with fixed barrier height \cite{dik1}, and the other member was the DW potential with variable barrier height but fixed positions of the two degenerate minima. The oscillons production in these two members of the family of DK DW potentials was established, and shown to occur when the distance between the minima gets smaller for the first member, and when the barrier height becomes lower for the second member. Based on their results with the two DK DW models, the authors concluded that the lowering of kink energy with increase of the shape deformability parameter was the determinant factor favoring the production of oscillons in the two models.\par
Stimulated by the studies of Bazeia {\it et al} \cite{13,31}, in the present work we investigate kink-antikink collisions and the possible production of oscillons in a DW model with fixed barrier height and fixed separation between the two degenerate minima, but a variable curvature of the barrier hump. With this we wish to establish that parametrized DW models with increasing kink energy as a function of a deformability parameter, are also quite prone to production of oscillons upon kink-antikink collisions. In fact we will show that the anharmonicity of the potential at its maximum, characterized by a flat barrier hump with increasing deformability parameter, is more likely to represent the unifying factor favoring the production of oscillons in the DK DW hierarchy. Proceeding with we shall introduce a new member to the family of DK DW potentials \cite{dik2}, characterized by a parametrization that leaves unaffected the barrier height and positions of the two potential minima, but allows tuning the steepness (or the curvatures) of the potential walls causing the barrier hump to flatten out. \par
In Section \ref{sec2} we introduce the member of parametrized DK DW potential with variable steepness, and formulate its field-theoretical dynamics. This enables us determine some associate characteristic quantities such as its kink and antikink solutions and the kink creation energy. In section \ref{sec3} we examine the kink-antikink scatterings, with emphasis on the production of oscillons as the deformability paramater is varied. Section \ref{sec4} is devoted to a summary of results and to conclusion.

\section{The model, kink solution and kink-phonon scattering spectrum}
\label{sec2}
\begin{figure}\centering
\includegraphics[width=3.in,height=2.in]{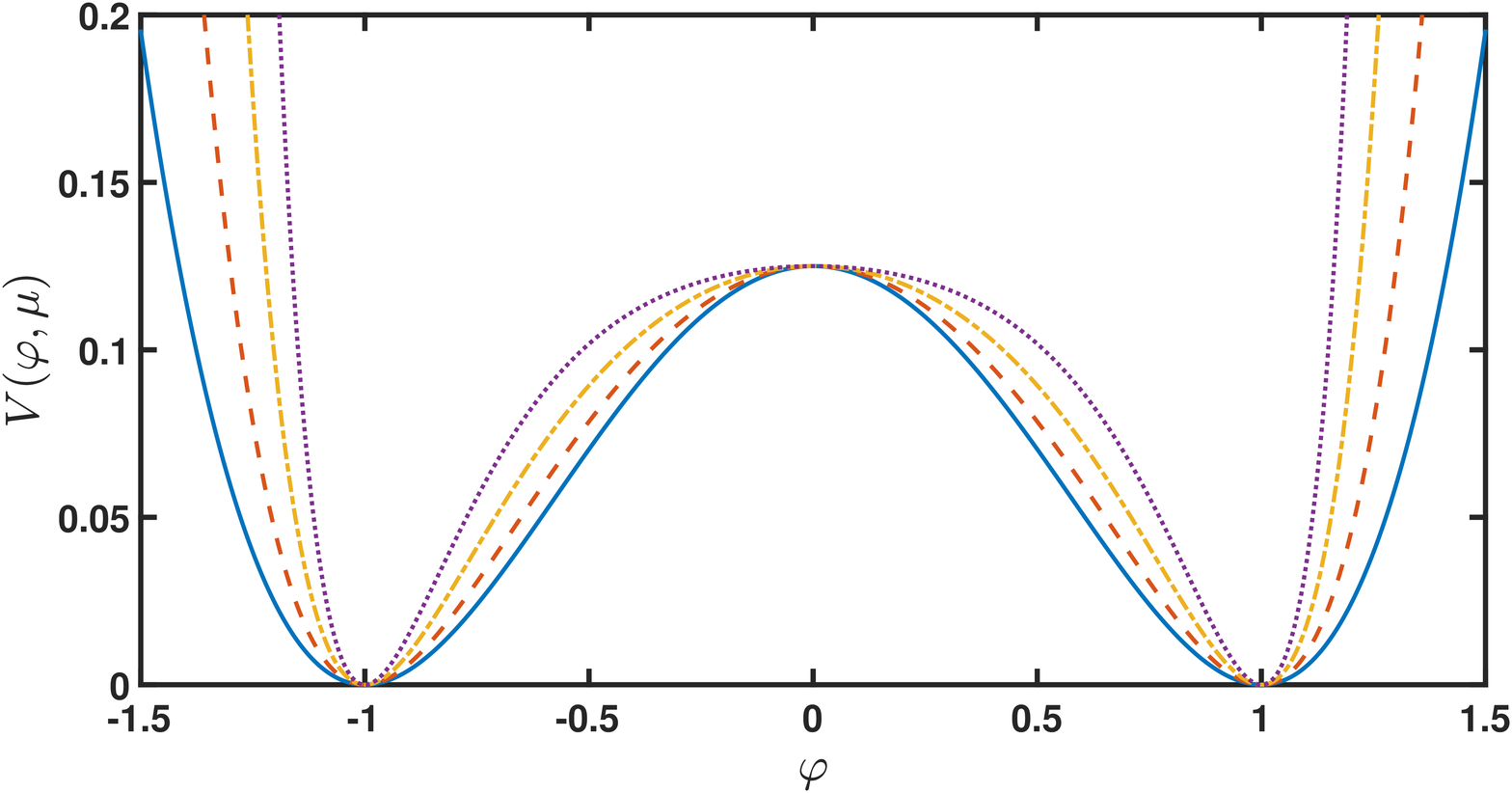}
\caption{(Color online) Plot of the double-well potential $V(\varphi,\mu)$, for some values of $\mu$:  $\mu= 0$ (Solid line), $\mu=2.0$ (Dashed line), $\mu=4.0$ (Dot-dashed line),  $\mu=8.0$ (Dotted line).} 
\label{fig1}
\end{figure}

Consider a field-theoretical model in ($1+1$) dimensional space-time, the dynamics of which is described by the Lagrangian:
\begin{equation}
L = \frac{1}{2}\left( \frac{\partial \varphi}{\partial t} \right)^{2} - \frac{1}{2}\left(\frac{\partial \varphi}{\partial x}\right)^{2}  - V(\varphi,\mu),
\label{e1}
\end{equation}
where $\varphi(x,t)$ is a real scalar field in one space ($x$) and temporal ($t$) dimensions. $V(\varphi,\mu)$ is a one-body scalar potential which can be expressed more generally \cite{dik2}:
\begin{equation}
V(\varphi,\mu) = \frac{1}{8} \left( \frac{\sinh^{2}(\alpha(\mu) \varphi)}{\mu^{2}} - 1\right)^{2},  \hskip 0.3truecm  \mu > 0.
\label{e2}
\end{equation}
In the present study we pick:
\begin{equation}
\alpha(\mu)= asinh(\mu), \label{con}
\end{equation}
which is a function of a real parameter $\mu$ assumed to control shape profile of the DW potential. For arbitrary values of the shape defomability parameter $\mu$, the scalar potential $V(\varphi,\mu)$ is a bistable function symmetric around a potential barrier located at the equilibrium state $\varphi=0$. The potential possesses two degenerate vacuum states at $\varphi = \pm 1$. Thus, unlike the two members of the DK DW potential discussed by Bazeia {\it et al} in ref. \cite{31}, the barrier height and minima positions of the parametrized DW potential (\ref{e2}) are always fix. However, on fig. \ref{fig1}, where $V(\varphi,\mu)$ is sketched for some values of $\mu$, one sees that the variation of $\mu$ influences the steepness of the potential walls. 
Quite interesting fig. \ref{fig1} suggests that the change in steepness of the potential walls, caused by a variation of the deformability parameter $\mu$, has the consequence of rendering the top of the potential barrier either flat or sharp. Indeed, when $\mu$ tends to zero the parametrized DW potential (\ref{e2}) reduces exactly to the standard $\phi^{4}$ potential \cite{cur,schrief}:
\begin{equation}
V(u)=\frac{1}{8}\left(u^2 - 1\right)^2.
\end{equation}
As $\mu$ increases the minima positions and the barrier height remain unchanged, but the slope of the potential walls gets steeper: the narrowest part of the potential barrier broadens while the flatness (i.e. the anharmonicity) of the barrier hump (or top) becomes more pronounced, resulting in an enhancement of the confinement of the two potential wells.

\begin{figure}\centering
\begin{minipage}[c]{0.32\textwidth}
\includegraphics[width=1.8in,height=1.5in]{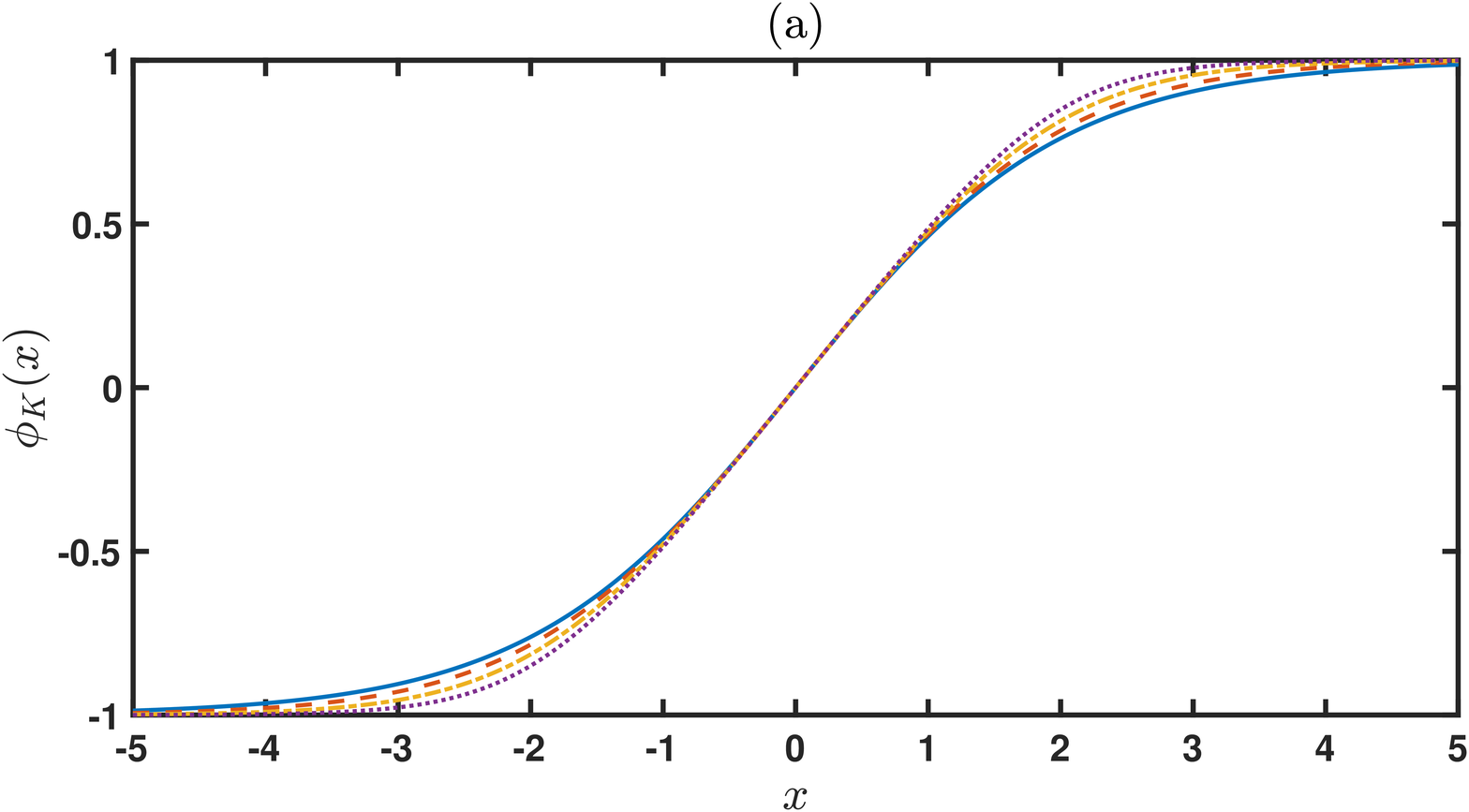}
\end{minipage}\hfill
\begin{minipage}[c]{0.32\textwidth}
\includegraphics[width=1.8in,height=1.5in]{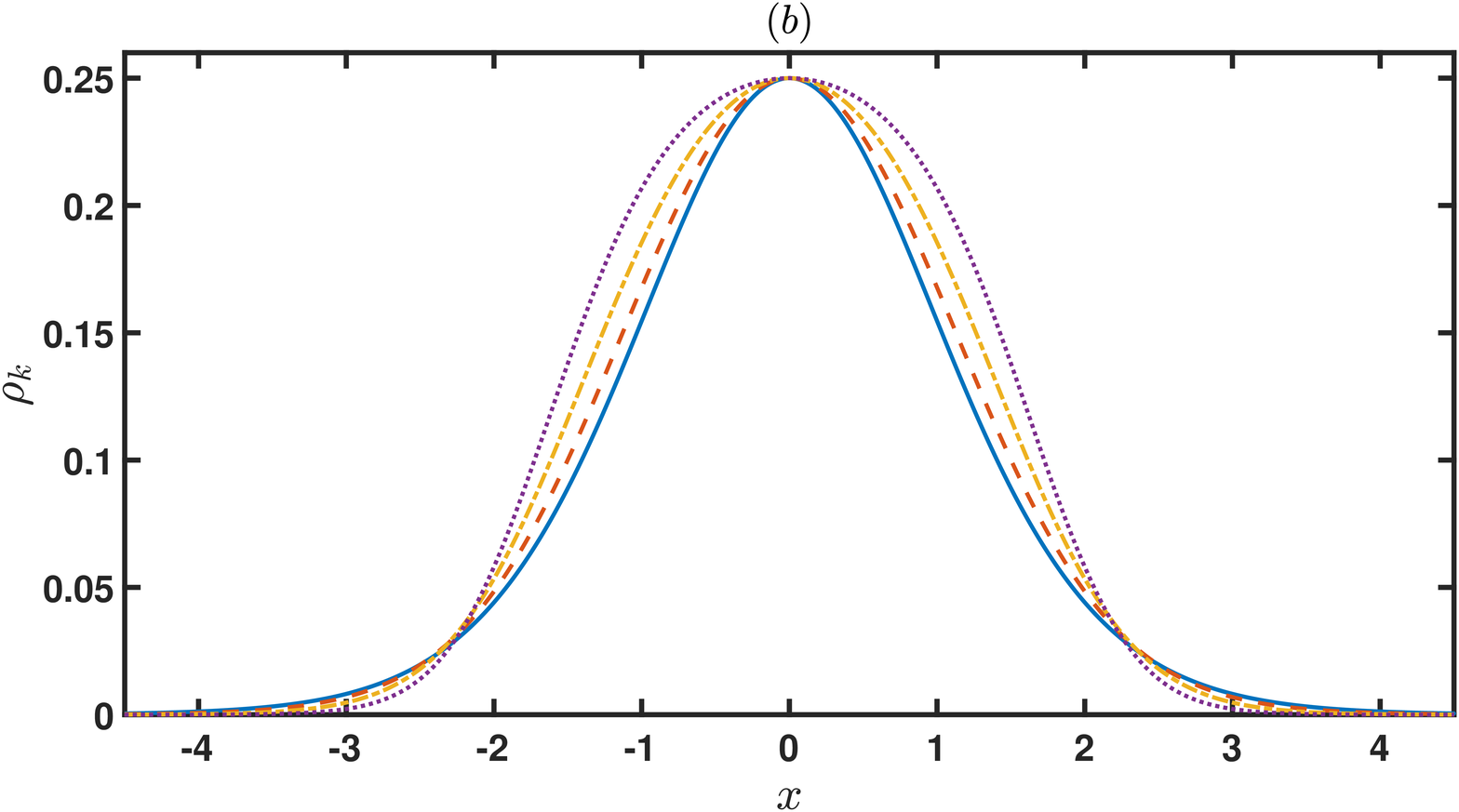}
\end{minipage}\hfill
\begin{minipage}[c]{0.32\textwidth}
\includegraphics[width=1.8in,height=1.5in]{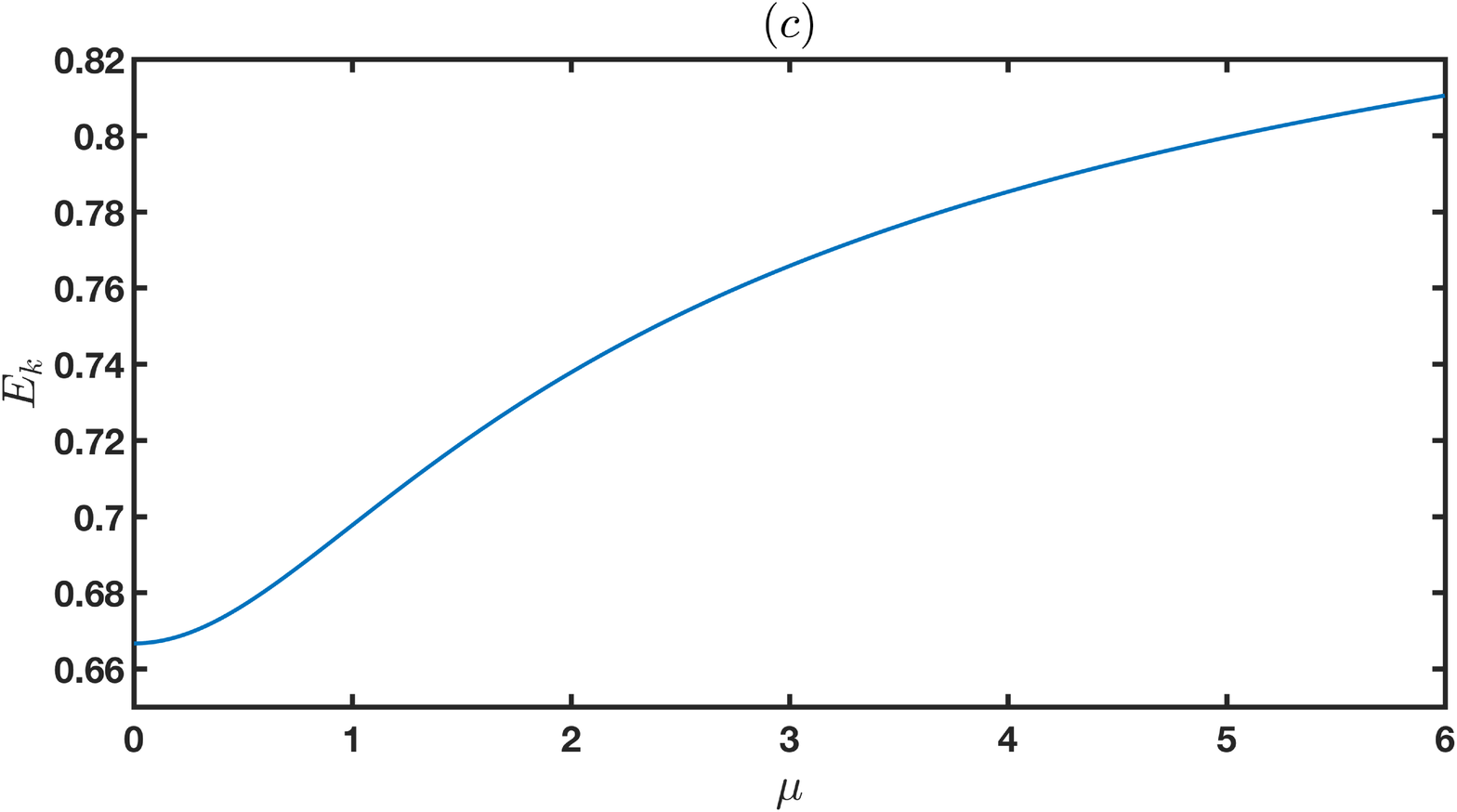}
\end{minipage}
\caption{(Color online) (a) Shape of the kink $\varphi_{K}(x)$ and (b) of the energy density $\rho_{\mu}(x)$ as a function of $x$, for:  $\mu= 0$ (Solid line), $\mu=2.0$ (Dashed line), $\mu=4.0$ (Dot-dashed line) and $\mu=8.0$ (Dotted line). (c) Variation of the kink creation energy $E_{K}$, as a function of $\mu$. } 
\label{fig2}
\end{figure}

The Lagrangian in formula (\ref{e1}) leads to the following equation of motion for the field $\varphi$:  
\begin{equation}
\frac{\partial^{2} \varphi}{\partial t^{2}} - \frac{\partial^{2} \varphi}{\partial x^{2}} + \frac{d}{d\varphi}V(\varphi,\mu)=0.
\label{e3}
\end{equation} 
In the static regime, the solitary-wave solution to this equation is given by:
\begin{equation}
\varphi_{K,\bar{K}}(x)= \pm \frac{1}{\alpha(\mu)}\tanh^{-1}\left[\frac{\mu}{\sqrt{1+\mu^{2}}} \tanh \frac{\sqrt{2}x}{d(\mu)}\right],
\label{e4}
\end{equation}
where: 
\begin{equation}
d(\mu)= \frac{2\mu}{\alpha(\mu)\sqrt{(1+\mu^{2})}}.
\label{e5}
\end{equation}
The solution with "+" sign stands for a kink $\varphi_{K}(x)$, while the solution with "-" sign stands for an antikink $\varphi_{\bar{K}}(x)$ of width $d(\mu)$. The characteristic energy (or rest mass) associated with the static kink and static antikink solution eq. (\ref{e4}), is obtained by using the general expression: 
\begin{equation}
E_K = \int^{+\infty}_{-\infty}\rho_{\mu}(x)dx, 
\label{e6}
\end{equation}
with: 
\begin{equation}
\rho_{\mu}(x) =  \frac{1}{2}\left(\frac{\partial \varphi}{\partial x}\right)^{2}  + V(\varphi,\mu)
\label{e7}
\end{equation}
the kink energy density. Substituting the solitary-wave solution obtained in formula (\ref{e4}) this yields:
\begin{equation}
E_{K}  =\frac{1}{4\alpha(\mu)\mu^{2}}\left[ 2\alpha(\mu)(1+\mu^{2}) - sinh(2\alpha(\mu))\right].
\label{e8}
\end{equation}
In fig. \ref{fig2}, shape profiles of the static kink solution $\varphi_{K}(x)$ (a) and of the kink energy density $\rho_{\mu}(x)$ (b), are plotted versus the spatial coordinate $x$ for some values of the deformability parameter $\mu$. The bottom graph in the figure, i.e. graph (c), represents the variation of the kink rest energy as a function of the deformability parameter $\mu$. 
One sees that as the deformability parameter $\mu$ increases, the asymptotic values of $\varphi_{K}(x)$ as $\vert x\vert\rightarrow\infty$ remains the same but a decrease in the kink width is noticeable. On the other hand, an increase of $\mu$ leaves the maximum of the energy density unaffected but affects the width of the energy density in the region covered by the barrier and the potential wells. Remarkably the energy density seems to decrease with $\mu$ as we go far in the region covered by the repulsive walls of the potential, such that in this region the kink is expected to become more localized. \par
Fig. \ref{fig2}c depicts the kink rest energy as a monotonically increasing function of the shape deformability parameter. In other words, an increase in $\mu$ will enhance the kink stability and hence the sharpness of the kink profile.\par
Most of the processes from the kink-antikink collisions arise as a consequence of vibrational modes inherent to the kink scattering excitation spectrum. Usually a perturbation theory is utilized to derive the spectrum of localized excitations around a kink \cite{33,cur}. To this last point, perturbing linearly the scalar field $\varphi(x,t)$ around the one kink solution $\varphi_{K}(x)$ i.e. $\varphi(x,t) = \varphi_{K}(x) + \eta(x)\exp(-i\omega t)$, yields the following Schr\"odinger-like eigenvalue problem \cite{33,cur}:
\begin{equation}
\left[-\frac{\partial^{2}}{\partial x^{2}} + V_{sch}(x, \mu)\right] \eta = \omega^{2}\eta.
\label{e9} 
\end{equation}
{\bf In this eigenvalue equation the quantity {\bf $V_{sch}(x, \mu) = \frac{d^{2}V}{d\varphi^{2}}\vert_{\phi_K}$} is the scattering potential, which in the present case is given by:}
\begin{eqnarray}
V_{sch}(x, \mu) &=& a_{0}\frac{\left[ \mu^{2}\tanh^{4}\left(\frac{x}{d(\mu)}\right) + 3\tanh^{2}\left(\frac{x}{d(\mu)}\right)-c(\mu)\right]}{\left[ d(\mu)\left(\mu^{2}\tanh^{2}\left(\frac{x}{d(\mu)}\right) - c(\mu)\right)\right]^{2}}, \nonumber \\ 
c(\mu) &=& 1+\mu^{2}. \label{e10} 
\end{eqnarray}
Note that this scattering potential determines the kink stability upon scattering with phonons \cite{33,cur}. Instructively an identical expression for $V_{sch}(x, \mu)$ is obtained by taking a linear perturbation around an antikink solution. \par
\begin{figure}\centering
\includegraphics[width=2.5in,height=2.in]{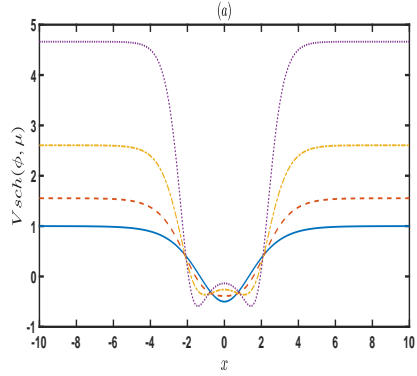}
\caption{(Color online) Plot of the scattering potential $V_{sch}(x, \mu)$ as a function of $x$, for $\mu=0$ (Solid line), $\mu=1.0$ (Dashed line), $\mu=2.0$ (Dot-dashed line) and $\mu=3.0$ (Dotted line).} 
\label{fig3a}
\end{figure}

 Distinct profiles of the scattering potential $V_{sch}(x, \mu)$, for different values of the deformability parameter $\mu$, are represented in fig. \ref{fig3a}. 
One sees that as $\mu$ increases, the scattering potential has its width that gradually decreases and its asymptotic limit growing drastically higher. In the range $ 0<\mu \lesssim 1.2$ the potential possesses a global minimum located at $x=0$, which transforms into a local maximum together with the appearance of two degenerate minima in the potential as the value of $\mu$ rises larger than $1.2$. \par
\begin{figure}\centering
\includegraphics[width=2.5in,height=2.in]{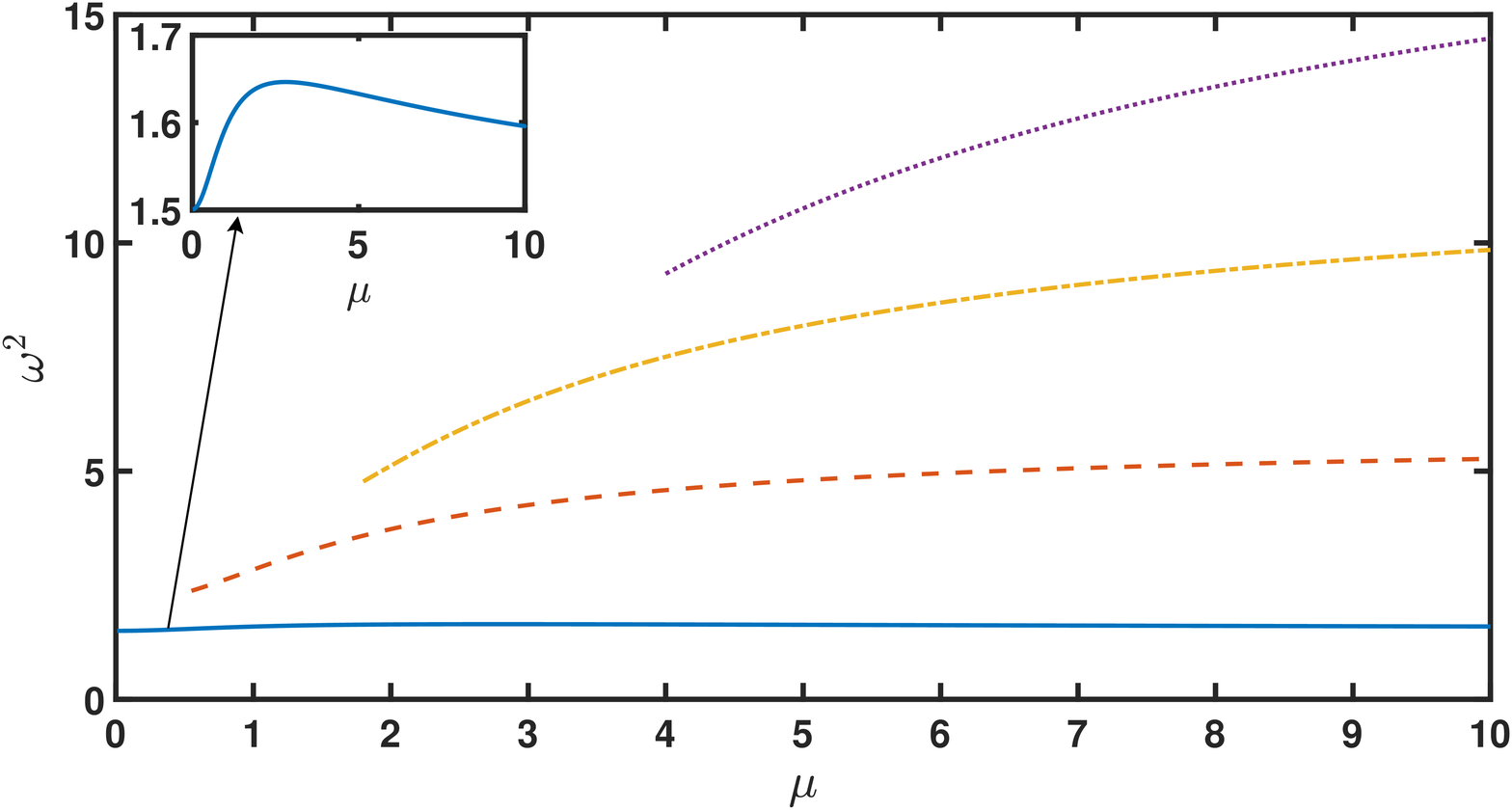}
\caption{(Color online) Plot of the squared frequencies $\omega^{2}$ of the vibrational states, as a function of $\mu$.} 
\label{fig3b}
\end{figure}

 
The same way as the scattering potential, the occurrence of bound states holds a key importance in grasping some relevant features of the scattering structure of the system \cite{13,31}. In particular a resonance mechanism for the exchange of energy between the translational mode and a vibrational mode, may result in rich consequences in the spectral features of the system \cite{31}. To gain insight onto this last featurte, we solved the eigenvalue equation (\ref{e9}) for $\mu$ and results emphasizing the influence of the parametrization on the appearance of bound states, are shown in fig. \ref{fig3b}. 
We note the presence of a zero-mode for all the values of the shape deformability parameter, moreover the appearance of new bound states is observed as $\mu$ rises. For instance, as $\mu$ lies in the range $0.55 \lesssim \mu \lesssim 1.8 $ we notice the presence of two vibrational states, and in the ranges $1.8 \lesssim\mu \lesssim 4.0 $ and $\mu \gtrsim 4 $ a third and a fourth bound state emerge respectively. Furthermore the lower vibrational has its frequency increasing as $\mu$ grows to a specific value, then decreasing while the frequencies of higher vibrational states are monotonically increasing functions of the deformability parameter.

\section{Analysis of kink-antikink collisions}
\label{sec3}
The dynamical equation pertaining to colliding kink-antikink pairs will be solved numerically in this section, with the aim to explore some characteristic spectral features of kink-antikink scatterings and identify vibrational models associated with the collisions. To this end, eq. (\ref{e3}) is discretized on a spacial grid with periodic boundary conditions. The grid is divided into $N$ nodes such that zone widths $\delta x$ in the simulations have fixed size, with the location of the $n^{th}$ point on the grid given by $x_{n}=n\Delta x$. The scalar field is then defined by $\varphi_{n}(t)=\varphi( x_{n},t)$ for $n = 1,2,...,N$. The second-order spatial derivative is approximated using a fourth-order central-difference scheme \cite{36}, which leads to a set of $N$ coupled second-order ordinary differential equations in $\varphi_{n}$ i.e.:
\begin{eqnarray}
\frac{\partial^{2} \varphi_{n}}{\partial t^{2}} = \frac{1}{12(\Delta x)^{2}}(-\varphi_{n-2} + 16\varphi_{n-1}-30\varphi_{n} \nonumber \\ +16\varphi_{n+1}  
- \varphi_{n+2}) - \frac{dV(\varphi_{n},\mu)}{d\varphi_{n}} ,
\label{e11} 
\end{eqnarray}
which is solved numerically using a fourth-order runge-kutta scheme with fixed step. The accuracy of our algorithm stands with errors that scale as $(\Delta x)^{2}$ and $(\Delta t)^{4}$.\par
The initial data used in our simulations represent a kink and antikink centered at the points $x=-x_{0}$ and $x=x_{0}$ respectively, and moving forward each other with initial velocities $\upsilon$ in the laboratory frame. The definition of the starting function can therefore be expresssed:
\begin{eqnarray}
\varphi(x,0) = \varphi_{K}(x+x_{0}, v, 0) - \varphi_{K}(x-x_{0}, -v, 0) - \varphi_{m}, \nonumber \\
\label{e12} 
\end{eqnarray}
{\bf where $\varphi_{m}=\pm 1$ for the kink-antikink and the antikink-kink initial configurations, respectively}. We set the grid to be sufficiently large, with left and right boundaries respectively at $x_{l}=-400$ and $x_{r}=+400$, and the separation distance to be $2x_{0}= 24$. The choice of a large grid together with periodic boundary conditions, was to avoid the reflected kink forms to travel to the boundary, and also to prevent any radiation emitted during the collision process to eventually find itself back to interact with the kinks. The grid is discretized with $N = 10^{5}$ nodes and all simulations were run with a temporal step size $\Delta t = 0.7(\Delta x)$, found to be a good consensus between the costly computational time and the production of results form high-resolution runs.\par
\begin{figure*}\centering
\includegraphics[width=5.6in,height=3.in]{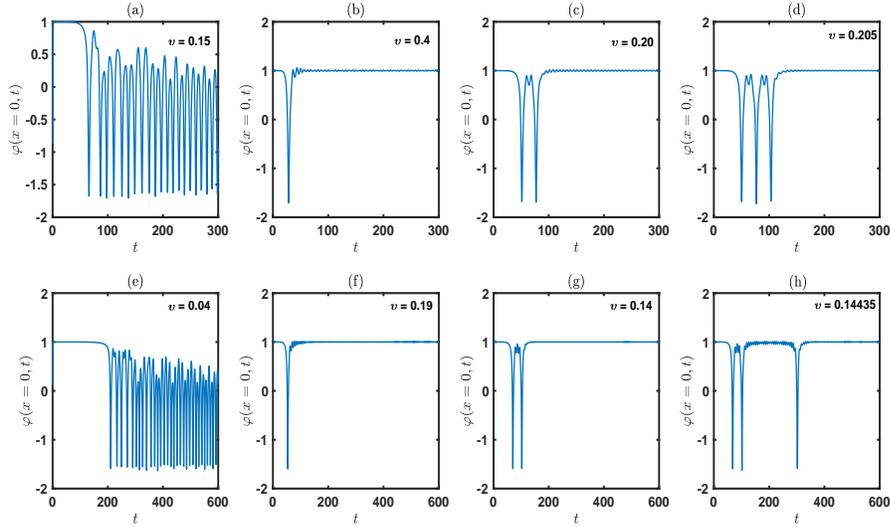}
\caption{(Color online) Possible results for a kink-antikink collision at several initial velocities, considering two values of the shape deformability parameter $\mu$. $\mu= 0$: (a), (b),(c) and (d). $\mu = 0.5$: (e), (f),(g) and (h). Values of $\mu$ were choosen such that only one vibrational mode sppears in the excitation spectrum.} 
\label{fig4}
\end{figure*}
\begin{figure}\centering
\includegraphics[width=2.5in,height=2.in]{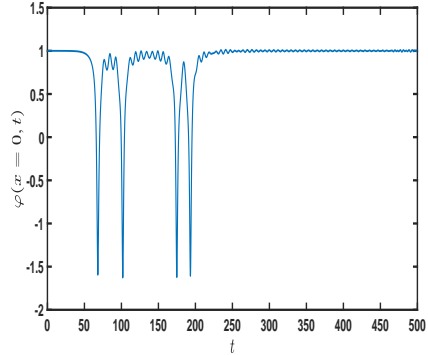}
\caption{(Color online) A four-bounce window taking $\mu = 0.5$ is evidenced by plotting $\varphi(x=0,t)$  for $\upsilon = 0.1445$. Note the presence of four large spikes illustrating collision after which the kink and antikink reflect and recede from each other forming a two-soliton state.} 
\label{fig5}
\end{figure}
\begin{figure}\centering
\begin{minipage}[c]{0.5\textwidth}
\includegraphics[width=2.6in,height=2.2in]{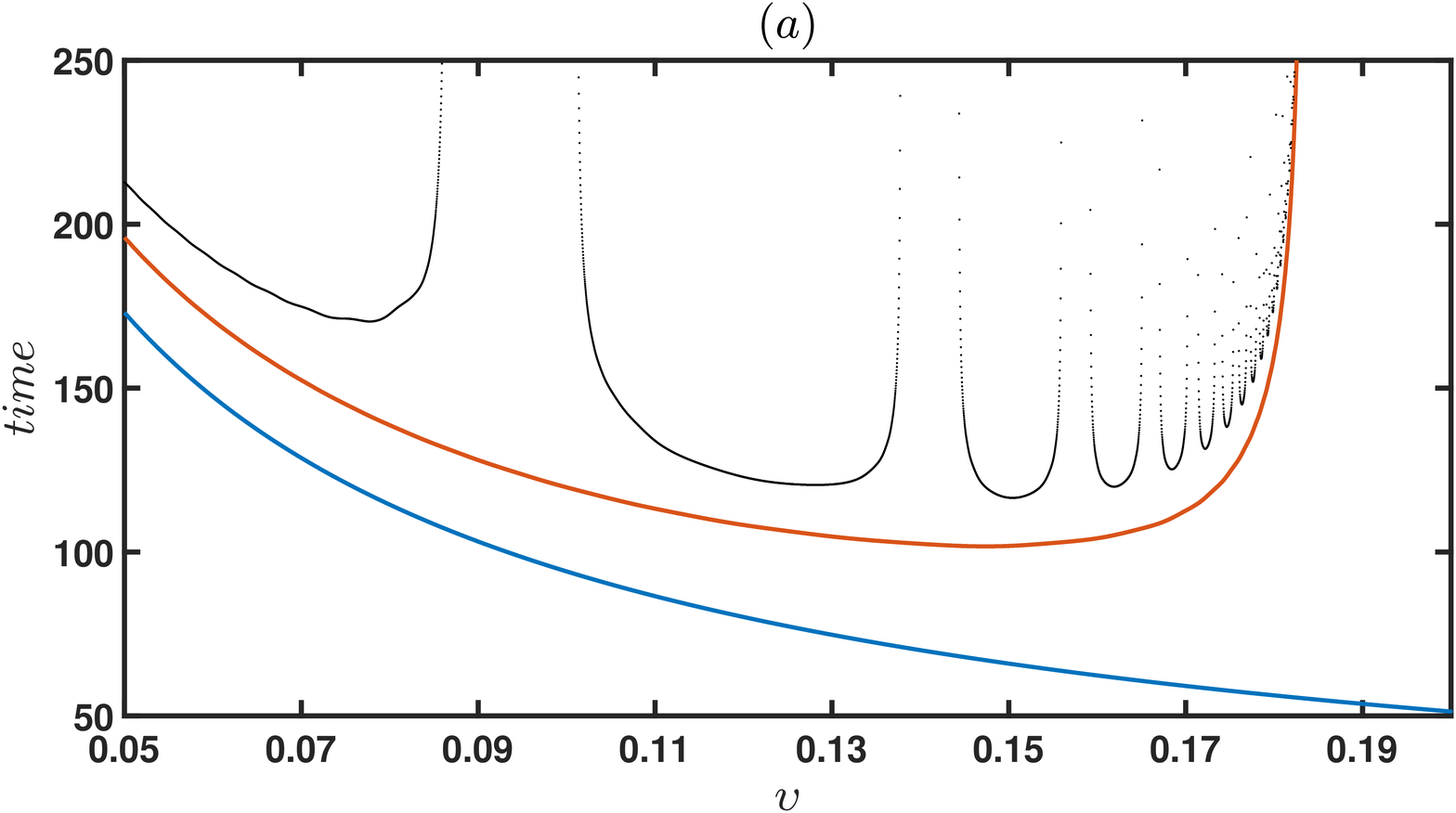}
\end{minipage}\hfill
\begin{minipage}[c]{0.5\textwidth}
\includegraphics[width=2.6in,height=2.2in]{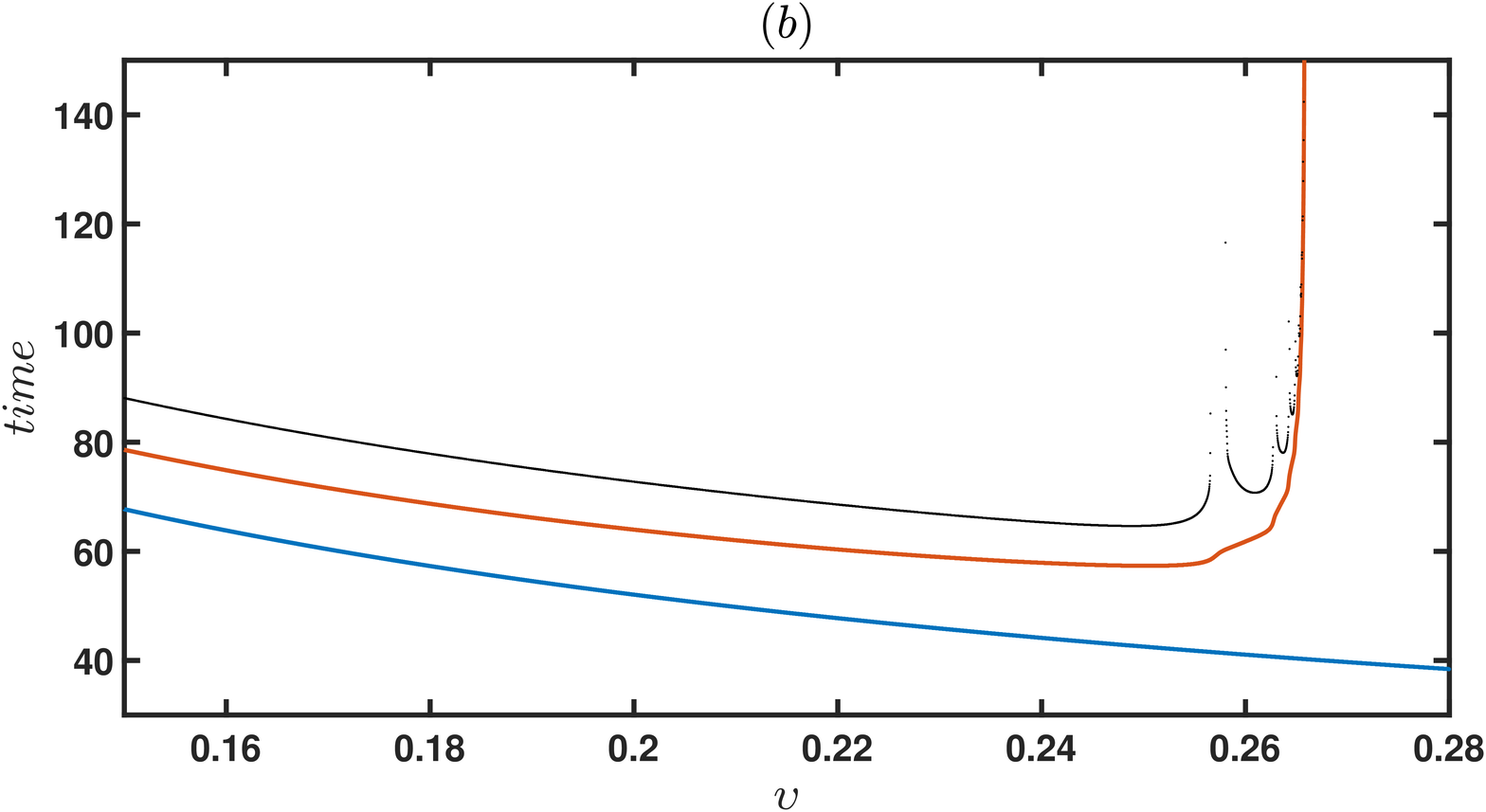}
\end{minipage}
\caption{(Color online) Kink-antikink collision times to first (blue), second (red) and third (dotted-black) bounces, as a function of initial velocity for (a) $\mu = 0.5$ and (b) $\mu = 5.0$. } 
\label{fig6}
\end{figure}

Several outputs obtained from numerical simulations at some different initial velocities are now doscussed. For weak velocities, the kink and antikink are expected to bound upon collision and have enough time to radiate sufficient energy forming a bion state. This is shown in figs. \ref{fig4}(a) and \ref{fig4}(e), where we plotted the evolution of the center of mass $\varphi(x=0,t)$ of the kink-antikink pair, for some values of the shape deformability parameter for which the system has just one vibrational mode. Irrespective of the barrier deformation, the kink-antikink pair moving with an initial velocity lower than a critical velocity $\upsilon_{c}$ settles to an erratically oscillating bion state. But for large velocities $\upsilon>\upsilon_{c}$, the kink-antikink pair does not have enough time to radiate sufficient energy to form a bion state. Thus the kink and antikink will once collide and permanently reflect each other during the scattering process. This is represented in figs. \ref{fig4}(b) and \ref{fig4}(f).
For all the considered values of $\mu$ one can note the appearance of a spike illustrating the collision, followed by a leveling off at $\varphi = +1$ implying that the kink and antikink have reflected and traveled far from each other. Still, the transition between the bion state and the reflection state is not smooth as the initial velocities increase. There are regions of values of $\upsilon$ for which these two states alternate. This regions were reported in several works as "windows" \cite{1e,2,14,16,31}. For instance, in figs. \ref{fig4}(c) and \ref{fig4}(g), we note two spikes in the evolution of the center of mass implying that the kink and antikink collide and reflect, then return to collide again before receding to a permanent reflection. Referring to the number of collisions before the last and permanent reflection, the sets of contiguous initial velocities leading to this state can be identified as forming a two-bounce window. The presence of a three-bounce window is evidenced in figs. \ref{fig4}(d) and \ref{fig4}(h), the velocities lying in the three-bounce windows are found on the edge of the two-bounce regions. When values of the deformability parameter $\mu$ are located in the range where there exists only one vibrational mode, the system presents a similar fractal structure as the one observed from the scattering process in the $\phi^{4}$ model. For example we can note the existence of a four-bounce window in fig. \ref{fig5}.
The appearance of n-bounce windows is also expected to be observed for this range of shape deformability parameter values. To further understand the structure of scattering in the system, in fig. \ref{fig6} we plotted the time of the three bounces as a function of the initial velocity.
The intervals in which the time for the third collisions diverges, range in the two-bounce windows. A case with only one vibrational mode is illustrated in Fig. \ref{fig6}(a), which shows plot of the collision times for $\mu = 0.5$ (compare with figs. \ref{fig4}(e), \ref{fig4}(f), \ref{fig4}(g) and \ref{fig4}(h)). We can note that bion states are formed for $\upsilon<\upsilon_{c}\sim0.183$ while for $\upsilon>\upsilon_{c}$, the collision results in an inelastic scattering between the pair corresponding to one-bounce around one vacuum. Moreover, the figure captures the complete set of two-bounce windows of which the width continuously decreases and accumulates around $\upsilon_{c}$. The scattering times in the presence of several vibrational modes are plotted in fig. \ref{fig6}(b) where we considered the shape of the potential for $\mu=5$. We first see that $\upsilon_{c}$ grows larger with $\mu$. From our simulations we observed that the variation of the critical velocities is not a monotonic function of the shape deformability parameter, we first observed a decrease of $\upsilon_{c}$ as $\mu$ increases till $\mu\sim 1$ then recedes to an increasing behavior (this result is not presented here). This reflects that the attraction between the kink and antikink lessens as the system is deformed departing from the $\phi^{4}$ model, but the attractive interaction gets stronger and stronger as  $\mu\gtrsim1$.  Suppression of two-bounce windows is also observed in fig. \ref{fig6}(b). This is justified by the presence of several vibrational modes complicating the energy transfer from the translational mode to just one vibrational mode to achieve resonance conditions. \par
\begin{figure}\centering
\begin{minipage}[c]{0.5\textwidth} 
\includegraphics[width=2.6in,height=2.6in]{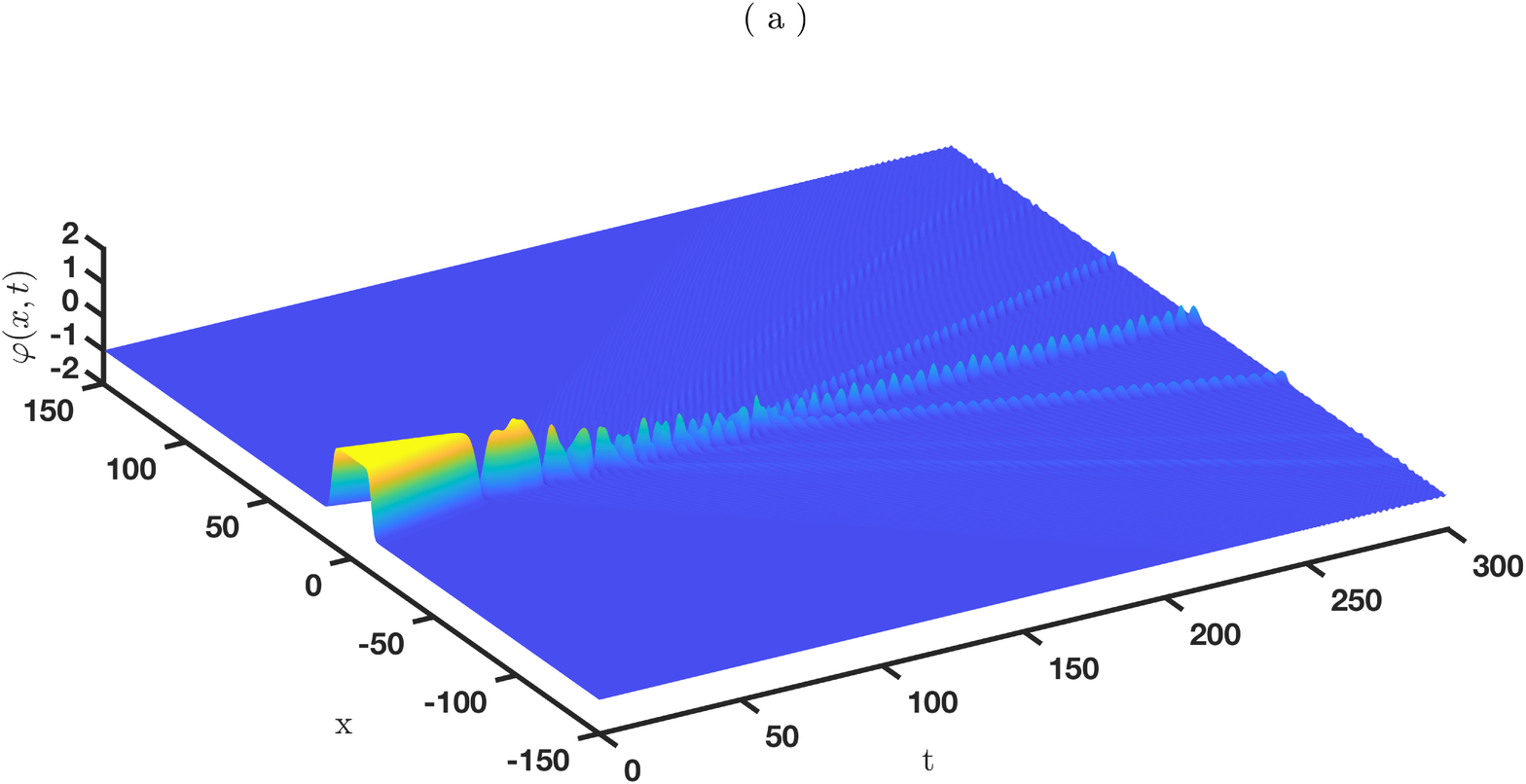}
\end{minipage}\hfill
\begin{minipage}[c]{0.5\textwidth}
\includegraphics[width=2.6in,height=2.6in]{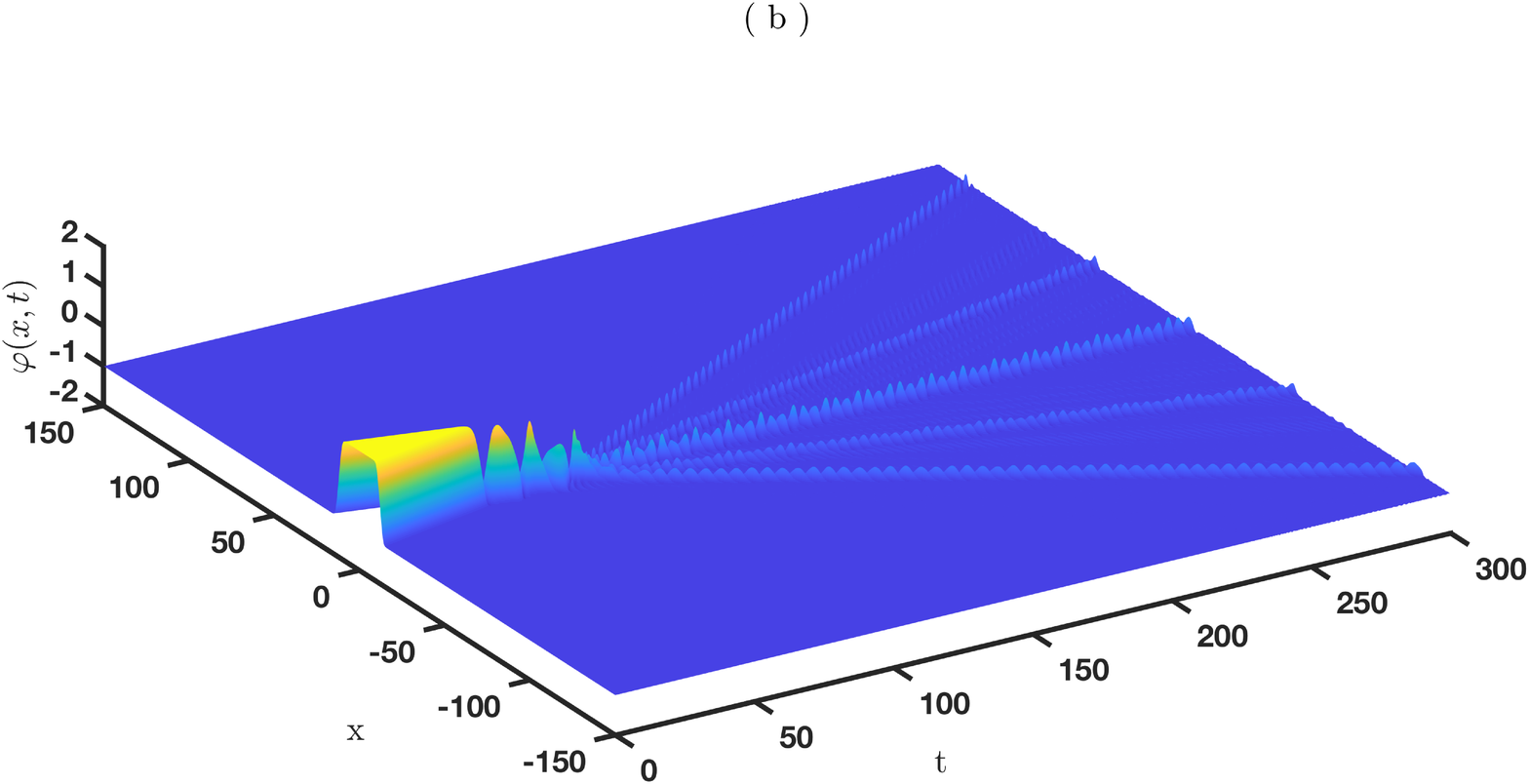}
\end{minipage}
\caption{(Color online) Oscillons resulting from kink-antikink collisions: (a) $\mu = 3.0$ and $\upsilon = 0.23$ $ (\upsilon_{c} = 0.245)$ showing one bion and two oscillons, (b) $\mu = 5.0$ and $ \upsilon = 0.236$ $(\upsilon_{c} = 0.27)$ showing one bion and four oscillons.} 
\label{fig7}
\end{figure}

One of our main point of focus in this work is the possible production of oscillons. For relatively small values of the deformability parameter $\mu$, for which only one vibrational mode is generated in the excitation spectrum, oscillon structures cannot appear, the kink-antikink collisions with initial velocities lower than the critical velocity can only result in bion and n-bounce states. The $\phi^{4}$ model being an asymptotic limit of the parametrized DW model in this range of values of $\mu$, this agrees with the $\phi^{4}$ model showing no evidence for the formation of oscillons as a result of the scattering process. We see from fig. \ref{fig7} that for larger values of $\mu$, the presence of more than one vibrational mode favors the production of oscillons.
At some velocities of the colliding kink, lower than the critical velocities, the bion is formed and travels with oscillons which can oscillate around each other, or escape to infinities. We can compare the appearance of one bion and two oscillons travelling together with a large flux of emitted radiation in fig. \ref{fig7}(a), and the appearance of one bion and four oscillons in fig. \ref{fig7}(b), travelling with quite no radiation and a larger degree of harmonicity. Note that the larger the shape deformability parameter the greater the number of created oscillons. 

\section{Conclusion }
\label{sec4}
Oscillons are breather-like bound states generated by self-interactions of kink-antikink pairs that exist in some scalar-field models \cite{jan,jan1,jan2,jan3,adam}, in the context of cosmology their built-in mechanism suggests that they can affect the standard picture of scalar ultra-light dark matter. In two recent studies \cite{13,31} the generation of oscillons in bistable systems, characterized by a parametrized double-well potential, was discussed with emphasis on the influence of the shape deformability on the oscillon production. First \cite{13} the authors considered a deformable $\phi^4$ potential represented by an hyperbolic double-well potential, and established that the deformability favors the emergence of oscillon modes from kink-ankink collisions and for well selected intial velocities of the colliding kinks. Later on \cite{31} they extended the study to two members of the family of Dikand\'e-Kofan\'e DW potentials. One of these members has its double-well minima fixed but a variable height of the potential barrier, whereas the other member has fixed barrier height but a variable separation between the two potential minima. \par 
To determine more exactly which of the characteristic features introduced by the potential deformability-i.e. the variable positions of potential minima, or the variable height of the potential barrier- effectively controls the oscillon production, in this work we revisited the study by considering a parametrized DW potential with fixed potential minima and fixed barrier height fixed. However the steepness of the potential walls, and hence the flatness of the barrier top, can be tuned by varying a deformability parameter. The parametrized DW potential has the particularity to reduce to a $\phi^{4}$ potential, just the same as with the already known family of DW potentials proposed in refs. \cite{dik1,dik2,33} and referred to as Dikand\'e-Kofan\'e potential. Examining the kink-antikink scattering processes, we found that the parametrized bistable model inherits some of the general features of the $\phi^{4}$ model that is the possibility of formation of bion states, reflected states and also $n-$bounce windows. However the appearance of additional modes in the scattering spectrum, as the DW potential deformation becomes predominant in our model, suggests the possibility of suppression of the two-bounce windows due to a kind of interference, as was already detailed in some other works \cite{16,31}.\par
Long-lived, quasi-harmonic and low-amplitude structures called oscillons were shown to form after kink-antikink collisions with some initial velocities less than a critical velocity. This is not observed for low values of $\mu$, where the model has only one vibrational state and is more close to the $\phi^4$ model. The rising number of vibrational states as $\mu$ increases yields to an intricate situation where the realization of the mechanism of resonant energy exchange between the translational and one vibrational more becomes more difficult. The appearance of oscillons is thus favored by the deformation in our model. \par 
In the works of Bazeia {\it et al}, reporting the appearance of oscillons in hyperbolic models  \cite{31} for two deformable double-well potentials, they showed that the production of oscillons is boosted by applying the conformational changes from those potentials' deformability such as reducing the distance between the minima keeping the barrier height fixed, or decreasing the barrier height while keeping the minima fixed. They pointed out that the factor unifying the two contexts is the lowering of the kink energy by the deformability in the two models. The scattering dynamics at the center of mass in our present model are roughly the same as the one in the work of ref. \cite{13,31}, however the increase of the kink energy in our model disagrees with a tentative idea to extend the consideration of kink energy being a determinant unifying factor to a more general case. The deformation in our model is manifest through an increase of the steepness of the potential walls, with the barrier top becoming flattened hence imposing an anharmonic shape to the potential barrier. This trend can also be observed as an implicit result of the deformation in the two models considered by Bazeia {\it et al} \cite{31}, and also in the sinh-deformed $\phi^{4}$ model \cite{13} shown to also allow the creation of oscillons. Bistable systems modeled by potentials with anharmonic barrier are thus suggested to be good candidates to observe the formation of oscillons in kink-scattering processes.

\section*{acknowledgements}
the work of A. M. Dikand\'e is supported by the Alexander von Humboldt (AvH) Foundation.

\end{document}